\documentclass[aps,prl,10pt,preprint]{revtex4-1}
\usepackage{graphicx,color}
\usepackage{dcolumn}
\usepackage{bm}
\usepackage{amsmath,amsthm,amssymb}
\usepackage{braket,here}
\usepackage[rel]{overpic}
\usepackage{pict2e}

\newcommand{\ccc}[1]{%
\ifx11
#1%
\fi}

\begin{document}
\ccc{
\title{
General Formalism for Magnetic Anisotropy Constants
}

\author{Daisuke Miura}
\affiliation{%
Department of Applied Physics, Tohoku University, Sendai, 980-8579, Japan}
\email{dmiura@solid.apph.tohoku.ac.jp}

\author{Ryo Sasaki}
\affiliation{Department of Applied Physics, Tohoku University, Sendai, 980-8579, Japan}

\author{Akimasa Sakuma}
\affiliation{Department of Applied Physics, Tohoku University, Sendai, 980-8579, Japan}


\date{\today}

\begin{abstract}
Direct expressions for the magnetic anisotropy constants are given at a finite temperature from microscopic viewpoints.
In the present derivation, it is assumed that the Hamiltonian is a linear function with respect to the magnetization direction.
We discuss in detail the first-order constant $K_1$ and show that the results reproduce previous results.
We also apply our method to Nd$_2$Fe$_{14}$B compounds
and demonstrate that the temperature dependencies of the magnetocrystalline anisotropy constants $K_1$, $K_2$, and $K_3$ are successfully computed.
\end{abstract}


\maketitle
}
\def\HC{{H_\mathrm{c}}}
\def\E{{\mathrm{e}}}
\def\I{{\mathrm{i}}}
\def\D{{\mathrm{d}}}
\def\V{{\hat{\mathcal{V}}}}
\def\P{{\hat{\mathcal{P}}_n}}
\def\CEF{{\hat{\mathcal{H}}}_\mathrm{CEF}}
\def\CN#1{{C_n^{(#1)}}}
\def\HEX{{H_\mathrm{ex}}}
\def\tr{{\mathrm{Tr}\ }}
\def\hc{{\mathrm{h.c.}}}
\def\cc{{\mathrm{c.c.}}}
\def\abs#1{{\left|#1\right|}}
\def\ret{\notag\\&\qquad}
\def\nr{\notag\\}

Understanding and controlling magnetic anisotropy is a central issue in developing magnetic devices such as spin RAMs, high-density storage, and permanent magnets.
The magnetic anisotropy at room temperature (or above) is important at a practical level and should be evaluated in consideration of finite-temperature effects for materials whose magnetic properties strongly depend on temperature.
This letter presents appropriate expressions for computing temperature-dependent magnetic anisotropy constants (MACs) and physically interpreting them.

An early study of the temperature dependence of the magnetocrystalline anisotropy was first performed by Zener~\cite{Zener1954}, which was followed by Callen and Callen~\cite{Callen1963,Callen1966}.
Considering an insulating ferromagnet having localized moments,
they phenomenologically derived a very simple expression, the so-called Callen--Callen law:
$
K_n(T)=K_n(0)\left[M(T)/M(0)\right]^{n(2n+1)},
$
where $K_n(T)$ is the $n$th-order MAC at the temperature $T$, and $M$ is the magnetization.
This theory is based on the concept in which a decrease in the MAC originates from the breakdown of the asphericity of the electron cloud of the magnetic ions due to thermal excitation~\cite{Skomski1999}.
Because a spherical electron cloud means the absence of a specific direction for the magnetic polarization (note that this does not mean the disappearance of the local moment),
the temperature dependence of the MAC can be related to that of the magnetization, as shown above.
After this pioneering work, there have been few theoretical works on the temperature dependence of the MACs.
Much later, Skomski \textit{et al}.~\cite{Skomski2006,Skomski2009,Skomski2013} discussed the Callen--Callen law by analyzing various magnetic materials and showed
that it can be applied to simple systems such as Fe and Co.
However, they also pointed out
that complex magnetic materials do not obey this law, such as NdCo$_5$ and Nd$_2$Fe$_{14}$B, which exhibit a spin reorientation.
Thus far, the Callen--Callen theory is not always satisfactory quantitatively for the range of applicability.

In recent years, theoretical trials have been conducted to evaluate the magnetocrystalline anisotropy energy (MAE) at a finite temperature
for spin Hamiltonians including the magnetic anisotropy term by a Monte Carlo approach~\cite{Evans2014,Matsumoto2015}.
Further, from a technological viewpoint, micromagnetic simulations of the Landau--Lifshitz--Gilbert equation including the random field describing thermal noise~\cite{Evans2014,Nakatani1997,Palacios1998,Nishino2015}
or the Landau--Lifshitz--Bloch equation~\cite{Garanin1997,Garanin2004} have been performed to observe the thermal stabilization of the magnetization.

Using a first-principles approach, the MACs at $T=0$ of transition-metal systems and rare-earth (RE) compounds have been calculated,
and the results were at a satisfactory level compared with the experimental data within the numerical accuracy.
For transition-metal systems, considerable effort has been made to clarify the effects of practical factors such as the interfaces~\cite{Wang1994,Daalderop1994} and distortion~\cite{Kyuno1992}
in magnetic multilayers and the chemical disorder~\cite{Aas2011,Kota2012} in alloys on the MAE.
For the finite-temperature treatment within the first-principles approach, Staunton \textit{et al}.~\cite{Staunton2004} and Matsumoto \textit{et al}.~\cite{Matsumoto2014}
have successfully demonstrated the temperature dependencies of the MAE of FePt and YCo$_5$, respectively,
on the basis of the functional integral method.
As for RE compounds, the MACs at $T=0$ have been successfully evaluated by calculating the crystalline electric field (CEF) acting on the 4f electrons
in the RE ions using the first-principles approach~\cite{Richter1992,Fahnle1993,Moriya2009}.
In the 1970s, Wijn \textit{et al}.~\cite{Wijn1973} and Buschow \textit{et al}.~\cite{Buschow1974} calculated the magnetic stabilization energy
related to a Sm ion in SmCo$_5$ from the Helmholtz free energy using ligand field theory.
Recently, we have theoretically studied the temperature dependence of $K_1$ and $K_2$ of Nd$_2$Fe$_{14}$B and well-reproduced the experimental data for $K_i(T)$
using appropriate crystalline field parameters~\cite{Sasaki2015}.
The method employed in this work is to derive $K_1$ and $K_2$ numerically from the free energy calculated as a function of the magnetization direction.

All approaches mentioned above have been rapidly developed in the past two decades owing to the considerable progress in numerical techniques and computer performance.
However, it is still difficult to measure $K_i$ directly, especially at a finite temperature,
because of the lack of an explicit quantitative expression for $K_i$.
Given this background, we provide a general expression for $K_i$ in the present work,
starting from the Hamiltonian describing the magnetic system, and describe a method to evaluate $K_i$ quantitatively for realistic ferromagnetic systems.
In particular, for localized electronic systems such as RE compounds
or systems that can only be described by angular momentum operators,
we demonstrate that the proposed method is useful and convenient for evaluating the temperature dependence of $K_i(T)$.

\def\H{{\hat{\mathcal{H}}}}
\def\tr{{\mathrm{Tr}}}
First, we derive the microscopic expression for the first-order MAC, $K_1(T)$, of crystals with $N$-fold rotational symmetry ($N\ge 3$).
Introducing the Helmholtz free energy, $F(\theta,\phi):=-\beta^{-1}\ln\sum_n\exp[-\beta E_n(\theta,\phi)]$
and comparing $F(\theta,\phi)$ with the phenomenological expression~\cite{Yamada1988,Herbst1991},
we have the relation
\ccc{
\begin{align}
K_1(T)&=\frac{1}{2}\left.\partial_\theta^2 F(\theta,\phi)\right|_{\theta=0},\label{eq:K1}
\end{align}
}
where $\beta:=1/(k_\mathrm{B}T)$ is the inverse temperature corresponding to $T$,
and $E_n(\theta,\phi)$ is an energy eigenvalue of the Hamiltonian $\H(\theta,\phi)$.
Let us consider cases where the Hamiltonian is explicitly written as a function of the polar angles $\theta$ and $\phi$
denoting the direction of the magnetization relative to the $c$ axis (the hat denotes an operator):
\ccc{
\begin{align}
\H(\theta,\phi):=\hat h+\bm m(\theta,\phi)\cdot\hat{\bm D},
\end{align}
}
where
$
\bm m(\theta,\phi):=(\sin\theta\cos\phi,\sin\theta\sin\phi,\cos\theta),
$
and $\hat h$ is the angle-independent part. $\hat{\bm D}$ is an operator with the transformation property of
\ccc{
\begin{align}
\hat C_N^\dagger\hat D^\pm \hat C_N&=\E^{\pm\I\frac{2\pi}{N}}\hat D^\pm,\label{eq:sym1}\\
\hat C_N^\dagger\hat D^z\hat C_N&=\hat D^z,\label{eq:sym2}
\end{align}
}
where $\hat D^\pm:=\hat D^x\pm\I\hat D^y$, and $\hat C_N$ is the $N$-fold rotational symmetry operator around the symmetrical axis parallel to the $c$ axis.
The derivatives of $F(\theta,\phi)$ are expressed as
\ccc{
\begin{align}
\partial_\theta F
&=
\sum_n\E^{-\beta \left(E_n-F\right)}
\partial_\theta E_n,
\label{eq:F1}
\\
\partial^2_\theta F
&=
\sum_n\E^{-\beta \left(E_n-F\right)}
\left[
\partial_\theta^2 E_n
-
\beta\left(\partial_\theta E_n\right)^2
\right]
+
\beta
\left(
\partial_\theta F
\right)^2,
\label{eq:F2}
\end{align}
}
in terms of the derivatives of $E_n(\theta,\phi)$.
As above in the purely theoretical treatment, the MACs can be described in terms of $E_n(\theta,\phi)$ in the neighborhood of $\theta= 0$.
Introducing $\V(\theta,\phi):=\bm m(\theta,\phi)\cdot\hat{\bm D}-\hat D^z$,
we divide the Hamiltonian $\H(\theta,\phi)$ into the unperturbative part $\H_0:=\hat h+\hat D^z$ and the perturbative part $\V(\theta,\phi)$.
Here, we notice that $\V(\theta,\phi)\to 0$ in the limit $\theta\to 0$ and
\ccc{
\begin{align}
\left[\hat C_N,\H_0\right]=0.\label{eq:sym3}
\end{align}
}
To obtain $K_1$, we perturbatively calculate $E_n$ up to the second order in $\V$.
We assume that the unperturbed eigenvalues $\epsilon_n$ and eigenstates $\ket{n}$ are given, which satisfy
the unperturbed Schr\"odinger equation, $\H_0\ket{n}=\epsilon_n\ket{n}$.
Assuming that the values of $\epsilon_n$ are not degenerate,
the perturbative energy terms are expressed as
$
\Delta_n^{(1)}=\braket{n|\V|n},
\Delta_n^{(2)}=\braket{n|\V\P\V|n}
$,
where $\P:=\sum_{k\neq n}\ket{k}\bra{k}/(\epsilon_n-\epsilon_k)$, and we have let $E_n(\theta,\phi)\simeq\epsilon_n+\Delta_n^{(1)}(\theta,\phi)+\Delta_n^{(2)}(\theta,\phi)$.
Considering the symmetries in Eqs.~(\ref{eq:sym1}) and (\ref{eq:sym3}), we obtain $\braket{n|\hat D^+|n}=\braket{n|\hat D^+\P\hat D^+|n}=0$
such that
$
\left.\partial_\theta E_n(\theta,\phi)\right|_{\theta=0}
=0
$ and
$
\left.\partial_\theta^2 E_n(\theta,\phi)\right|_{\theta=0}
=
-\braket{n|\hat D^z|n}
+
[\braket{n|\hat D^+\P\hat D^-|n}+\braket{n|\hat D^-\P\hat D^+|n}]/2
$.
Thus, from Eqs.~(\ref{eq:K1}), (\ref{eq:F1}), and (\ref{eq:F2}), we obtain
\ccc{
\begin{align}
&K_1(T)
=\frac{1}{2}\sum_n\E^{-\beta(\epsilon_n-F_0)}
\left[
-\braket{n|\hat D^z|n}
+
\frac{1}{2}\braket{n|(\hat D^+\P\hat D^-+\hat D^-\P\hat D^+)|n}
\right],
\label{eq:r1}
\end{align}
}
where $F_0:=F(0,\phi)$. This is one of main results used to compute $K_1(T)$ later in this letter.
Before discussing this expression, let us further transform it.
If we find an operator $\hat{\bm J}$ satisfying
\ccc{
\begin{align}
[\hat J^\alpha,\hat J^\beta]=\I\epsilon_{\alpha\beta\gamma}\hat J^\gamma,\label{eq:J1}\\
[\hat J^\alpha,\hat D^\beta]=\I\epsilon_{\alpha\beta\gamma}\hat D^\gamma,\label{eq:J2}
\end{align}
}
we can then eliminate the matrix elements of $\hat D^\pm$ with the help of $\hat{\bm J}$.
Here, $\epsilon_{\alpha\beta\gamma}$ is the Levi--Civita tensor, the repeated Greek indices are summed for all cases ($\alpha=x,y,z$),
the condition in Eq.~(\ref{eq:J1}) means that $\hat{\bm J}$ is an angular momentum operator,
and the condition in Eq.~(\ref{eq:J2}) geometrically means that $\hat{\bm D}$ is a vector operator for $\hat{\bm J}$.
Using the identity $\braket{n|\hat D^\pm|k}=\pm(\epsilon_n-\epsilon_k)\braket{n|\hat J^\pm|k}\pm\braket{n|[\hat J^\pm,\hat h]|k}$,
we obtain the commutator form
\ccc{
\begin{align}
&K_1(T)
=-\frac{1}{8}\braket{[\hat J^-,[\hat J^+,\hat h]+\hc]}
+\frac{1}{4}
\sum_n\E^{-\beta(\epsilon_n-F_0)}
\sum_{k\neq n}
\frac{|\braket{n|[\hat J^+,\hat h]|k}|^2+|\braket{k|[\hat J^+,\hat h]|n}|^2}{\epsilon_n-\epsilon_k},
\label{eq:r2}
\end{align}
}
where $\braket{\cdots}$ denotes the statistical average in $\H_0$.
This is another one of the main results that is appropriate for describing physical pictures of the magnetic anisotropy
because the commutators distill the essential features from the Hamiltonian.
Equations (\ref{eq:r1}) and (\ref{eq:r2}) are equivalent and general for $K_1$.

For example, let us apply these formulae to localized spin systems.
We consider the single-spin Hamiltonian
\def\SPIN{{\mathcal{H}_\mathrm{spin}}}
\def\LL{{\hat{\mathcal{H}}_\mathrm{4f}}}
\def\CRY{{\hat{\mathcal{H}}_\mathrm{cry}}}
\def\EX{{\bm H_\mathrm{ex}}}
\def\HEX{{H_\mathrm{ex}}}
\ccc{
\begin{align}
\SPIN:=-A(\hat S^z)^2-2\EX\cdot\bm{\hat S},\label{eq:spin}
\end{align}
}
where $\hat{\bm S}$ is a spin operator; $A$ is the single-ion MAE; $\EX$ is the exchange field parallel to the magnetization direction, i.e., $\EX=\HEX\bm m$;
and we assume the condition $A<2\HEX$.
Hence, we find the correspondences
$
\hat{\bm D}\to-2\HEX\bm{\hat S}
$ and
$
\hat h\to-A(\hat S^z)^2
$.
Obviously, $\hat{\bm S}$ satisfies the conditions in Eqs.~(\ref{eq:J1}) and (\ref{eq:J2}) as $\hat{\bm J}$.
Then, the commutators are calculated as
$
[\hat S^+,-A(\hat S^z)^2]
=A(\hat S^z\hat S^++\hat S^+\hat S^z)
$ and
$
[\hat S^-,[\hat S^+,-A(\hat S^z)^2]]
=2A[\hat{\bm S}^2-3(\hat{S}^{z})^2]
$.
From Eq.~(\ref{eq:r2}),
\ccc{
\begin{align}
K_1^\mathrm{spin}(T)
&=
-\frac{A}{2}\left[S(S+1)-3\sum_{M=-S}^S\E^{-\beta(\epsilon_M^\mathrm{spin}-F_0^\mathrm{spin})}M^2\right]\nr
&+\frac{1}{4}\sum_{M=-S}^S\E^{-\beta(\epsilon_M^\mathrm{spin}-F_0^\mathrm{spin})}
\left(\frac{A^2(1-2M)^2(S+M)(S-M+1)}{A(1-2M)-2\HEX}\right.
\nr
&+\left.\frac{A^2(1+2M)^2(S-M)(S+M+1)}{A(1+2M)+2\HEX}\right),
\end{align}
}
where $\epsilon_M^\mathrm{spin}:=-AM^2-2\HEX M$ is an energy eigenvalue of $\hat{\mathcal{H}}^0_\mathrm{spin}:= -A(\hat S^z)^2-2\HEX\hat S^z$, and $F_0^\mathrm{spin}:=-\beta^{-1}\ln\sum_{M=-S}^S\exp(-\beta\epsilon_M^\mathrm{spin})$.
At zero temperature, we have
\ccc{
\begin{align}
K_1^\mathrm{spin}(0)=AS\left(S-\frac{1}{2}\right)\frac{1}{1+A(S-1/2)/\HEX}.
\end{align}
}
It is clear that this tends to $AS^2$ in the classical limit $S\to\infty$, as expected from Eq.~(\ref{eq:spin}).
Now, we refer to the relation between $K_1^\mathrm{spin}(T)/K_1^\mathrm{spin}(0)$ and $S(T)/S(0)$.
Here, the statistically averaged spin is defined by $S(T)=\sum_{M=-S}^S\exp\left[-\beta(\epsilon_M^\mathrm{spin}-F_0^\mathrm{spin})\right]M$.
The Callen--Callen law states that $K_1^\mathrm{spin}(T)/K_1^\mathrm{spin}(0)$ is $\left[S(T)/S(0)\right]^3$.
To consider this power law, we introduce a temperature-dependent power
\ccc{
\begin{align}
\alpha(T):=\frac{\ln [K_1^\mathrm{spin}(T)/K_1^\mathrm{spin}(0)]}{\ln [S(T)/S(0)]}.
\end{align}
}
The zero-temperature value is exactly obtained as
\ccc{
\begin{align}
\alpha(0)=\frac{6+A(2S-3)/\HEX}{2+A(2S-3)/\HEX}.
\end{align}\\
}
Thus, our formula supports the Callen--Callen law: $\alpha(0)\to 3$ in the limit $A/\HEX\to 0$.
However, we observe a deviation from the law in the classical limit because $\alpha(0)$ monotonically decreases to 1 with respect to $S$.
Thus, the Callen--Callen law is valid under the condition $AS/\HEX\ll 1$.
Figure (\ref{fig:KS}) shows the temperature dependencies of the quantities with $S=1$ and $A/\HEX=0.1$.
We observe that $\alpha(T)$ slightly deviates from the Callen--Callen law because of corrections from the higher-order terms with respect to $A/\HEX$.

The other sample is a localized 4f-electron system.
Let us consider a model in which 4f electrons are in ligand and exchange fields~\cite{Yamada1988,Wijn1973,Herbst1991,Buschow2005}:
\ccc{
\begin{align}
\LL:=\lambda\hat{\bm L}_\mathrm{f}\cdot\hat{\bm S}_\mathrm{f}+\CRY+2\EX\cdot\hat{\bm S}_\mathrm{f},
\end{align}
}
where $\CRY$ and $\EX$ represent, respectively, the CEF and exchange field acting on the electrons;
$\lambda$ is the strength of the spin--orbit interaction; and $\hat{\bm L}$ and $\hat{\bm S}$ are the total orbital and total spin angular momenta of the electrons, respectively.
Assuming that $\EX=\HEX\bm m$, the correspondences are
$\hat{\bm D}\to 2\HEX\hat{\bm S}_\mathrm{f}$ and
$
\hat h\to\lambda\hat{\bm L}_\mathrm{f}\cdot\hat{\bm S}_\mathrm{f}+\CRY
$.
Using $\hat{\bm J}\to\hat{\bm S}_\mathrm{f}$, the conditions in Eqs.~(\ref{eq:J1}) and (\ref{eq:J2}) are satisfied.
Therefore,
\ccc{
\begin{align}
K_1^\mathrm{f}(T)&=-\frac{\lambda}{4}\braket{\hat{\bm L}_\mathrm{f}\cdot\hat{\bm S}_\mathrm{f}+\hat{L}_\mathrm{f}^z\hat{S}_\mathrm{f}^z}^\mathrm{f}
+\frac{\lambda^2}{4}\sum_n\E^{-\beta(\epsilon_n^\mathrm{f}-F_0^\mathrm{f})}
\sum_{k\neq n}\frac{\abs{\braket{n|(\hat{L}_\mathrm{f}^+\hat{S}_\mathrm{f}^z-\hat{L}_\mathrm{f}^z\hat{S}_\mathrm{f}^+)|k}^\mathrm{f}}^2+(n\leftrightarrow k)}{\epsilon_n^\mathrm{f}-\epsilon_k^\mathrm{f}}.
\label{eq:fK1}
\end{align}
}
At first glance, we can understand that $K_1^\mathrm{f}$ vanishes in absence of the spin--orbit interaction.
Now, we purturbatively evaluate Eq.~(\ref{eq:fK1}) with respect to $\lambda$ to the second order under the assumption that the CEF has tetragonal symmetry.
The straightforward calculation gives
\ccc{
\begin{align}
K_1^\mathrm{f}(T)\simeq\frac{\lambda^2}{4}\sum_n\E^{-\beta(\varepsilon_n-f)}\sum_{k\neq n}\frac{4\varDelta L_{nk}^z-\varDelta L_{nk}^+-\varDelta L_{nk}^-}{\varepsilon_k-\varepsilon_n},
\end{align}
}
where $(\CRY+2\HEX\hat S_\mathrm{f}^z)\ket{u_n}=\varepsilon_n\ket{u_n}$, $f:=-\beta^{-1}\ln\sum_n\exp(-\beta\varepsilon_n)$,
and $\varDelta L^\alpha_{nk}:=|\braket{u_n|\hat{L}_\mathrm{f}^z\hat{S}_\mathrm{f}^\alpha|u_k}|^2-|\braket{u_n|\hat{L}_\mathrm{f}^x\hat{S}_\mathrm{f}^\alpha|u_k}|^2$.
This is a version of a localized spin system for one in itinerant electronic systems~\cite{Bruno1989,Wang1993,Laan1998,Kota2014}.
Here, it is noteworthy that we can also regard $\hat{\bm J}_\mathrm{f}:=\hat{\bm L}_\mathrm{f}+\hat{\bm S}_\mathrm{f}$ as $\hat{\bm J}$ because $[\hat{L}_\mathrm{f}^\alpha,\hat D^\beta]=0$.
This choice leads to another form
\ccc{
\begin{align}
&K_1^\mathrm{f}(T)
=-\frac{1}{8}\braket{[\hat L^-_\mathrm{f},[\hat L^+_\mathrm{f},\CRY]+\hc]}^\mathrm{f}
+\frac{1}{4}
\sum_n\E^{-\beta(\epsilon_n^\mathrm{f}-F_0^\mathrm{f})}
\sum_{k\neq n}
\frac{|\braket{n|[\hat L^+_\mathrm{f},\CRY]|k}^\mathrm{f}|^2+(n\leftrightarrow k)}{\epsilon_n^\mathrm{f}-\epsilon_k^\mathrm{f}},
\label{eq:LK1}
\end{align}
}
where we have used $[\hat{\bm J}_\mathrm{f},\lambda\hat{\bm L}_\mathrm{f}\cdot\hat{\bm S}_\mathrm{f}]=\bm 0$ and $[\hat{\bm S}_\mathrm{f},\CRY]=\bm 0$.
Thus, we can easily confirm that $K_1^\mathrm{f}(T)=0$ when $\CRY=0$.
The commutators can be calculated by representing $\CRY$ in terms of spherical tensor operators constructed from $\hat{\bm L}_\mathrm{f}$ on the basis of the Wigner--Eckart theorem.
For simplicity, let us consider the minimum case that
\ccc{
\begin{align}
\CRY={C}_2^0\left[3\hat{L}_\mathrm{f}^z{}^2-\hat{\bm L}_\mathrm{f}{}^2\right]
\quad\text{and}\quad
\HEX\to\infty
\end{align}\\
}
for $\lambda\to\infty$, where $C_2^0$ is a constant encoding a CEF.
The commutators are calculated as $[\hat L^-_\mathrm{f},[\hat L^+_\mathrm{f},\CRY]]=-C_2^0[2\bm{\hat L}_\mathrm{f}^2-6\hat L^z_\mathrm{f}{}^2]$, and
the second term in Eq.~(\ref{eq:LK1}) vanishes in the limit of a large exchange field such that
$
K_1^\mathrm{f}(T)=\frac{3}{2}C_2^0\braket{\hat{\bm L}^2_\mathrm{f}-3\hat L^z_\mathrm{f}{}^2}
$.
Then, replacing $\hat{\bm L}_\mathrm{f}$ by $\hat{\bm J}_\mathrm{f}$ on the basis of
the equivalent-operator technique~\cite{Gerloch1983,Judd1998} for $\lambda\to\infty$, we have
$
K_1^\mathrm{f}(T)=\frac{3}{2}B_2^0\braket{\hat{\bm J}_\mathrm{f}^2-3\hat J^z_\mathrm{f}{}^2}
$.
Therefore, for light RE elements, our formula reproduces
\ccc{
\begin{align}
K_1^\mathrm{f}(T)=-3J(J-1/2)B_2^0,
\end{align}
}
where $B_2^0$ is a CEF parameter, and $J$ is a total angular momentum~\cite{Yamada1988}.
Note that in the limit of $\HEX\to 0$, we observe that $K_1^\mathrm{f}\to 0$ because the first and second terms in Eq.~(\ref{eq:LK1}) cancel.

Finally, we demonstrate numerical calculations for the temperature-dependent MACs ($K_1, K_2$, and $K_3$) of Nd$_2$Fe$_{14}$B. 
Following the derivation of Eq.~(\ref{eq:r1}) and purturbatively calculating $E_n(\theta,\phi)$ up to the fourth order,
the microscopic expressions of $K_2$ and $K_3$ are obtained as
\ccc{
\begin{align}
K_2&=\frac{1}{4}\sum_n\E^{-\beta(\epsilon_n-F_0)}[-\HEX\CN{1}+4\HEX^2\CN{3}\nr
&+4\HEX^3(\CN{1}\CN{5}-\CN{4})
-4\HEX^4(\CN{2}\CN{5}-\CN{6})]\nr
&+\frac{\beta}{2}\left[
K_1{}^2-\sum_n\E^{-\beta(\epsilon_n-F_0)}
(
\HEX\CN{1}-\HEX^2\CN{2}^2
)
\right]\label{eq:K2}
,\\
K_3&=2\HEX^4\sum_n\E^{-\beta(\epsilon_n-F_0)}\CN{7},\label{eq:K3}
\end{align}
}
where we have defined the constants as
$\CN{1}:=\braket{n|\hat S_\mathrm{f}^z|n}$,
$\CN{2}:=\braket{n|(\hat S_\mathrm{f}^-\P\hat S_\mathrm{f}^+ + \hat S_\mathrm{f}^+\P\hat S_\mathrm{f}^-)|n}$,
$\CN{3}:=\braket{n|\hat S_\mathrm{f}^z\P\hat S_\mathrm{f}^z|n}$,
$\CN{4}:=\bra{n}(\hat S_\mathrm{f}^-\P\hat S_\mathrm{f}^z\P\hat S_\mathrm{f}^+ +\hat S_\mathrm{f}^+\P\hat S_\mathrm{f}^z\P\hat S_\mathrm{f}^-+2\hat S_\mathrm{f}^-\P\hat S_\mathrm{f}^+\P\hat S_\mathrm{f}^z+2\hat S_\mathrm{f}^+\P\hat S_\mathrm{f}^-\P\hat S_\mathrm{f}^z)\ket{n}$,
$\CN{5}:=\braket{n|(\hat S_\mathrm{f}^-\P{}^2\hat S_\mathrm{f}^++\hat S_\mathrm{f}^+\P{}^2\hat S_\mathrm{f}^-)|n}$,
$\CN{6}:=\bra{n}(\hat S_\mathrm{f}^+\P\hat S_\mathrm{f}^+\P\hat S_\mathrm{f}^-\P\hat S_\mathrm{f}^-+\hat S_\mathrm{f}^-\P\hat S_\mathrm{f}^-\P\hat S_\mathrm{f}^+\P\hat S_\mathrm{f}^+ + \hat S_\mathrm{f}^+\P\hat S_\mathrm{f}^-\P\hat S_\mathrm{f}^+\P\hat S_\mathrm{f}^-
+
\hat S_\mathrm{f}^-\P\hat S_\mathrm{f}^+\P\hat S_\mathrm{f}^-\P\hat S_\mathrm{f}^+
+
2
\hat S_\mathrm{f}^+\P\hat S_\mathrm{f}^-\P\hat S_\mathrm{f}^-\P\hat S_\mathrm{f}^+
)\ket{n}$,
and
$\CN{7}:=\bra{n}\hat S_\mathrm{f}^+\P\hat S_\mathrm{f}^+\P\hat S_\mathrm{f}^+\P\hat S_\mathrm{f}^+\ket{n}$.
Figure \ref{fig:Nd} shows the temperature dependencies of the MACs for Nd$_2$Fe$_{14}$B calculated using Eqs.~(\ref{eq:r1}), (\ref{eq:K2}), and (\ref{eq:K3}).
Here, we have not repeated the discussion of the comparison with experimental results (see Ref~\cite{Sasaki2015}).
See Ref.~\cite{Wijn1973,Yamada1988} for detailed computational methods for solving the eigenproblems.
The expressions have the following advantages.
(I) There are no calculation parameters.
In contrast, a mesh parameter is necessary for finite-difference calculations of Eq.~(\ref{eq:K1}).
Although there are integral techniques using Fourier series, the spherical harmonics expansion~\cite{Birss1974}, and other expansions in terms of various complete sets,
mesh parameters are also needed.
(II) The calculation diagonalizing the Hamiltonian matrix only needs to be performed once.
If finite-difference or integral methods are used,
the eigenvalues at the mesh points need to be computed.
(III) For materials with small MACs, high-precision results can be expected
because the cancellation of significant digits caused by finite-difference calculations is avoided.

In summary,
we have derived direct expressions for the MACs at a finite temperature from microscopic viewpoints.
It is only assumed that the Hamiltonian is a linear function with respect to the magnetization direction.
We have discussed in detail the first-order constant, $K_1$, and shown that the results reproduce previous results.
Furthermore, we have successfully demonstrated the calculation of the temperature-dependent $K_1$, $K_2$, and $K_3$ for Nd$_2$Fe$_{14}$B.

\ccc{
\begin{acknowledgments}
This work was supported by JST--CREST.
\end{acknowledgments}
}

\bibliographystyle{apsrev4-1}

\providecommand{\noopsort}[1]{}\providecommand{\singleletter}[1]{#1}%

\ccc{
\begin{figure}
\centering
\includegraphics[width=5cm]{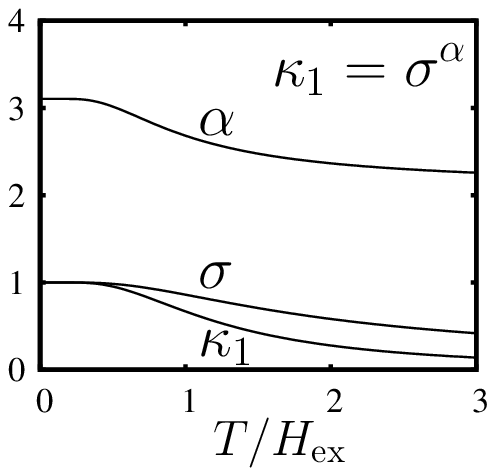}
\caption{Calculated results of $\sigma(T):=S(T)/S(0)$, $\kappa_1(T):=K_1(T)/K_1(0)$, and the power $\alpha(T)$ as a function of the temperature $T$ with $S=1$ and $A/\HEX=0.1$.
The Callen--Callen law predicts $\alpha(T)=3$.
}
\label{fig:KS}
\end{figure}

\begin{figure}
\centering
\includegraphics[width=6cm]{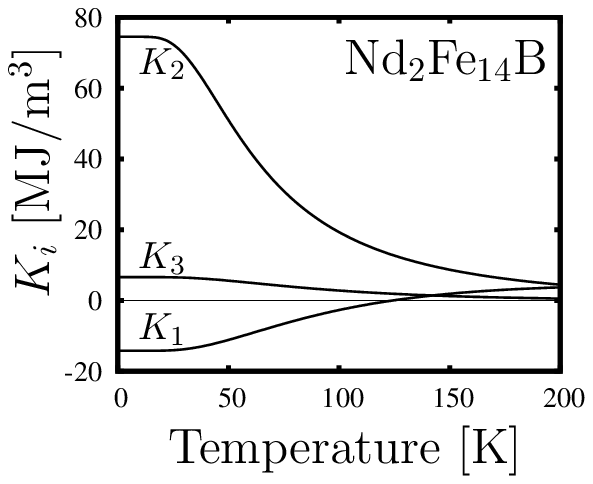}
\caption{Calculated magnetocrystalline anisotropy constants $K_1$, $K_2$, and $K_3$ as a function of temperature.
For Nd$_2$Fe$_{14}$B,
$\braket{\hat r^2}=1.001$ a$_0{}^2$,
$\braket{\hat r^4}=2.401$ a$_0{}^4$,
$\braket{\hat r^6}=12.396$ a$_0{}^6$,
$A_2^0=295$ K/a$_0{}^2$,
$A_4^0=-12.3$ K/a$_0{}^4$,
$A_6^0=-1.84$ K/a$_0{}^6$,
$A_6^4=-15.9$ K/a$_0{}^6$,
$\HEX=350$ K~\cite{Yamada1988},
$\lambda\to\infty$, 
and the lattice constants a $=$ b $=0.881$ nm, c $=1.221$ nm~\cite{Sagawa1987}.
}
\label{fig:Nd}
\end{figure}
}


\end{document}